\documentstyle[11pt]{article}
\newcommand{\Wt}{W_{,t} }
\newcommand{\Wtt}{ W_{,tt} }
\newcommand{\Wth}{ W_{,\theta} }
\newcommand{\Wthth}{ W_{,\theta\theta} }
\newcommand{\Pt}{ \Phi_{,t} }
\newcommand{\Ptt}{ \Phi_{,tt} }
\newcommand{\Pth}{\Phi_{,\theta} }
\newcommand{\Pthth}{\Phi_{,\theta\theta}  }
\newcommand{\reff}[1]{(\ref{#1})}
\newcommand{\dis}{\displaystyle}
\newtheorem{definition}{Definition}
\newtheorem{theorem}{Theorem}

\begin{document}

\begin{titlepage}

\begin{center}

% The Title
\Large{Asymptotic Behavior
      of the $T^3 \times R$ Gowdy Spacetimes} \\
\vspace{1.5 cm}

% The Author(s) Name(s)
\large{Boro Grubi\v{s}i\'{c}} \footnote{On leave from the Institute of
Theoretical Physics, University of Zagreb, Croatia}
\large{ and Vincent Moncrief}
\vspace {1 cm}

\normalsize{Department of Physics, Yale University\\
217 Prospect St. \\ New Haven, CT 06511\\}
\vspace {1 cm}

\begin{abstract}
We present new evidence in support of the Penrose's strong cosmic
censorship conjecture in the class of Gowdy spacetimes with $T^3$
spatial topology. Solving Einstein's equations perturbatively to all
orders we show that asymptotically close to
the boundary of the maximal Cauchy development the dominant term in
the expansion gives rise to curvature singularity for almost all
initial data. The dominant term, which we call the ``geodesic loop
solution'',
is a solution of the Einstein's equations
with all space derivatives dropped. We also describe the extent to
which our perturbative results can be rigorously justified.

\end{abstract}

\vfill
PAC Number: 0420

\end{center}
\end{titlepage}
\section{Introduction}

It is well known \cite{choquet-york} that in the class of globally
hyperbolic spacetimes arbitrary Cauchy data for the vacuum Einstein
equations have unique maximal developments (maximal Cauchy
developments). We would like to know under which conditions a similar
statement could be true in some larger class of spacetimes, possibly
all vacuum Einstein spacetimes satisfying certain differentiability
conditions (see \cite{chrusciel91} for a nice review). The
\mbox{Taub-NUT} spacetime \cite{misner-taub} is prototype of a
spacetime where the globally hyperbolic part (Taub) can be extended
into at least two \mbox{non-isometric} \mbox{Taub-NUT} spacetimes
\cite{chrusciel-isenberg}, thus violating the uniqueness of the
maximal development. Penrose's Strong Cosmic Censorship (SCC)
conjecture states that for almost all (generic) initial data the
maximal development is, indeed, globally hyperbolic, not larger, hence
restoring the uniqueness of the maximal development. The  spacetimes
like \mbox{Taub-NUT} obviously necessitate the exclusion of some
spacetimes as special (nongeneric). In the case of \mbox{Taub-NUT} it
is the high symmetry (its isometry group is four dimensional) that
makes it nongeneric.

To prove the SCC we would have to specify arbitrary Cauchy data, find
the maximal globally hyprebolic development, and show that for almost
all initial data that development is inextendible by proving, for
example, that:
\begin{enumerate}
\item
 The maximal globally hyprebolic development is geodesically
 complete or
\item
For every incomplete geodesics some curvature scalar blows up  when
approaching the incomplete end or ends (the singularity).
\end{enumerate}
The Hawking-Penrose singularity theorems \cite{hawking-ellis} tell us
that, generically, the first case is not true, but they do not tell us
anything about the behavior of the curvature when we approach the
incomplete end of a geodesic (the singularity). This is not surprising
since these theorems use Einstein's equations to a very limited
extent. To find the behavior of the curvature when approaching the
singularity we'll have to find the detailed asymptotic behavior of the
solutions of Einstein's equations close to the singularity.

The proof of the SCC for the general case is presently out of our
reach due to the complexity of Einstein's equations, but, to gain some
insight into the general case, we can try to prove the SCC in the
restricted class of spacetimes having an \mbox{$n$-dimensional}
isometry group, with least possible $n$. Following that strategy in
this paper we analyse the asymptotic behavior of the solutions of
Einstein's equations in the class of Gowdy spacetimes on $T^3 \times
R$; spacetimes that have \mbox{2-dimensional} $U(1) \times U(1)$
isometry group \cite{gowdy71,gowdy74}.

We develop a new method for studying the asymptotic, singular behavior
of large classes of cosmological solutions of Einstein's equations.
The method consists principally in showing that one can solve the
$n$-th order perturbed Einstein's equations for all $n$ using the
``variation of constants'' method, and then use the freedom in choice
of the $n$-th order solution to obtain an asymptotic (when approaching
the singularity) sequence of functions which is presumably the
asymptotic expansion of some exact solution of Einstein's equations.
This sequence is uniquely determined by the zeroth order solution
which we call the ``geodesic loop solution''. The zeroth order
equations are just the Einstein's field equations with ``space''
derivatives dropped and the method is applicable only to the
spacetimes whose dynamics is ``velocity-dominated'' in the sense of
Eardley, Liang and Sachs \cite{els72,im90}.

For the first two orders our method gives the same results as the
method suggested by Cosgrove, based on the ``multiple scales''
technique of applied analysis \cite{cosgrove}. The full extent of
applicability of our method is not yet known. In the present paper we
develop and apply it to the case of Gowdy metrics on $T^3 \times R$
whose asymptotic behavior has been analysed before by Mansfield, using
a different perturbative method \cite{mansfield89}. In a separate work
we shall apply it to $U(1)$ symmetric spacetimes. We believe, but this
has not yet been explicitly demonstrated, that our basic method should
be applicable to general, non-symmetric solutions of Einstein's
equations.

An advantage to starting with with the Gowdy metrics on $T^3 \times R$
is that much is known about their behavior. In particular on can prove
a ``global existence theorem'' to the effect that, in a suitably rigid
coordinate system (in which the time coordinate $\tau$ measures the
geometrical area of the two-tori which arise as orbits of the isometry
group), all sufficiently smooth Gowdy solutions extend globally to all
$\tau \in (0,\infty)$ and that this interval of existence exhausts the
maximal globally hyperbolic development \cite{moncrief81}. The same
reference  shows that the $\tau=$ const. hypersurfaces always approach
a ``crushing singularity'' of uniformly diverging mean curvature as
$\tau \rightarrow 0^+$ but approach a boundary of infinite three
volume as $\tau \rightarrow \infty$. The former boundary always arises
after a finite lapse of proper time ( as measured, say, along the
normal trajectories to the chosen foliation from some non-singular
reference Cauchy surface) making the spacetime past geodesically
incomplete, whereas the latter occurs only after an infinite lapse of
proper time. Our main interest here is  with the ``crushing singular''
boundary which occurs as $\tau \rightarrow 0^+$, we would like to show
that the spacetime is inextendible beyond that boundary . It is
expected to be curvature singular generically, but may in some special
cases instead correspond to Cauchy horizon across which (analytic)
extensions are possible \cite{moncrief81b,moncrief82,moncrief84}.

The mathematical methods used in proving the main results of reference
\cite{moncrief81} give only rather weak information about the
behaviors of the solutions near their crushing  boundaries at
$\tau=0$. For the special case of polarized Gowdy metrics (which are
governed by a linear hyperbolic equation) on can strengthen the
arguments of Ref.\cite{moncrief81} to derive genuinely sharp estimates
for the behaviors of the metric functions and their derivatives near
the crushing boundaries \cite{im90}. In this case on can rigorously
characterize the asymptotic behavior of every solution and even
classify the solutions in a natural way by studying the asymptotic
behaviors of their curvature tensors. While we have considerable hope
that similar arguments can be developed for the general, non-linear
Gowdy metrics, this has not yet been accomplished.

Lacking a direct (i.e., non-perturbative) means for establishing the
asymptotic, singular behavior of the generic Gowdy spacetime, we have
developed the perturbative approach to be presented here. This
perturbative approach has (as we have mentioned) the advantage of
applicability far beyond the rather limited class of Gowdy metrics.
When restricted to the Gowdy class however, it yields results which
may eventually be provable by direct, non-perturbative methods. For
the larger classes of spacetimes to which the perturbative method
definitely applies (e.g., those having one spacelike Killing field),
the possibility of a direct (non-perturbative) proof of the
corresponding results seems currently to be rather remote. Thus the
Gowdy metrics provide a natural test-case for the perturbative method.
The perturbative results provide natural conjectures which one can
reasonably hope to prove by direct methods and the direct methods, if
successful, can be expected to justify the (less rigorously founded)
perturbative approach.

The outline of the rest of the paper is as follows: In section 2 we
define the Gowdy spacetimes and briefly review the field equations. In
section 3.1 we formulate the perturbative method, in section 3.2 we
solve the zeroth and first order equations and find the asymptotic
behavior of the corresponding solutions. In section 3.3  we show that,
as $\tau\rightarrow 0^+$, all higher order solutions decay faster than
the zeroth order solution, called the  ``geodesic loop solution''. In
section 3.4 we show that for almost all initial data the ``geodesic
loop spacetimes'', which presumably capture the asymptotic behaviors
of the exact solutions of the Gowdy field equations, are
curvature-singular when, $\tau \rightarrow 0^+$, and hence
inextendible beyond their maximal Cauchy developments. In section 4 we
describe known rigorous results, which all support our perturbative
analysis.

\section{Gowdy $T^3 \times  R$ Spacetimes}

Consider a spacetime that can be foliated by a family of compact,
connected and orientable spacelike  hypersurfaces. If the maximal
isometry group of the spacetime is two-dimensional and if it acts
invariantly and effectively on the foliation than the isometry group
must be $U(1)\times U(1)$. Moreover, the foliation surfaces must be
homeomorphic to $T^3$ , $S^1 \times S^2$, $S^3$ or to a manifold
covered by one of these, and the action of the $U(1)\times U(1)$ is
unique up to a diffeomorphism. In such a spacetime the Killing vector
fields $K_a$, $a=1,2$ associated with the isometry group must commute,
and two scalar functions
\begin{equation}
c_a= \epsilon_{\alpha\beta\gamma\delta} K^{\alpha}_1 K^{\beta}_2
\nabla^{\gamma} K^{\delta}_a
\end{equation}
must be constant \cite{geroch72}. The spacetimes which satisfy
the above
mentioned symmetry requirements, and in which both constants $c_a$
vanish, will be called Gowdy spacetimes \cite{gowdy74}.

For the simplest case of $T^3$ spatial topology, Gowdy \cite{gowdy74}
showed that in such a spacetime a coordinate system can be found such
that the metric can be expressed in the form
\begin{eqnarray}
ds^{2} & = & e^{2 A} [ - e^{-2 t} dt^2 + d\theta^2 ]
          +  e^{- t} [ \cosh W + \cos \Phi \; \sinh W ] (dx^1)^2  \\
       & + & 2 e^{- t} \sin \Phi \;  \sinh W \;  dx dy
          + e^{- t} [ \cosh W - \cos \Phi \;  \sinh W ] (dx^2)^2 ,
\nonumber
\label{gowdymetric}
\end{eqnarray}
where $\partial / \partial x^1 $ and $\partial /\partial x^2$ are
commuting Killing vector fields on $T^3$ in the $S^1 \times S^1$
directions, and the functions $W$ ,  $\Phi$ and $A$ depend only on the
two remaining coordinates: ``time'' $t$ and ``angle'' $\theta$
parametrizing the third $S^1$ factor in $T^3$. For convenience,
instead of the time $\tau$ that measures the geometric area of the
two-dimensional group orbits we use the time $t=-\ln \tau$. The
singularity at $\tau=0$, mentioned in the introduction, corresponds to
$t\rightarrow \infty$.

Einstein's vacuum field equations give the equations of motion for the
metric functions $W$, $\Phi$ and $A$
\begin{eqnarray}
 \Wtt - \frac{1}{2}  \sinh W \Pt^2 =
   e^{- 2 t} [ \Wthth - \frac{1}{2}  \sinh 2W \; \Pth^2 ]  \label{i}\\
 \Ptt + 2 \coth W \; \Wt \Pt  =
  e^{- 2 t} [ \Pthth + 2 \coth W \; \Wth  \Pth ]  \label{ii} \\
 A_{,t} = \frac{1}{4}  [ 1 - \Wt^2 - \sinh^2 W \; \Pt^2  -   e^{-2t}
                                  ( \Wth^2 + \sinh^2 W \; \Pth^2 )]
\label{at} \\
 A_{,\theta}  =   - \frac{1}{2}  [ \Wt \Wth  + \Pt \Pth \sinh^2 W]
\label{atheta}
\end{eqnarray}
which have to be solved on the $t\theta$ cylinder.

To solve this system of equations we proceed in two steps. We first
solve equations \reff{i} and \reff{ii} for  $W$ and $\Phi$. These two
equations can be put into Hamiltonian form with the Hamiltonian
\begin{eqnarray}
H[X,\Pi] & = & \int_0^{2 \pi} {\cal H} d \theta  \\
{\cal H} &  = &  \frac{1}{2}
[ g^{AB} \Pi_A \Pi_B + e^{-2t} g_{AB} X^{A'}
                             X^{B'}  ] , \label{hamiltonian}
\end{eqnarray}
where $X^1$ and $X^2$ are $W$ and $\Phi$ respectively; prime denotes
differentiation with respect to $\theta$ and $g_{AB}$ is the metric of
two-dimensional hyperbolic space
\begin{equation}
 g= dW^2 +\sinh^2 W d\Phi^2.    \label{hyperbolic}
\end{equation}
{}From Hamilton's equations it follows that imposing periodicity in
$\theta$ (with period 2$\pi$) on initial conditions is enough to
obtain $W$ and $\Phi$ that are periodic at all times. A solution
$W(t,\theta)$ and $\Phi(t,\theta)$ can be conveniently pictured as a
closed loop moving in the two-dimensional hyperbolic space with speed
\begin{equation}
u= \sqrt{g^{AB} \Pi_A \Pi_B}.
\label{speed}
\end{equation}

After obtaining $W$ and $\Phi$  we can find $A$ by
evaluating the line integral
\begin{equation}
 A(t,\theta) = a_0 + \int_{\Gamma} \alpha
\label{intA}
\end{equation}
over some  path $\Gamma$ on  the $t \theta$ cylinder with the end
points at $(t_0,\theta_0)$ and $(t,\theta)$, and with components
$\alpha_t$ and $\alpha_{\theta}$ of the two dimensional one-form
$\alpha$ equal to the right-hand sides of the equations \reff{at} and
\reff{atheta} respectively. Expressed in Hamiltonian variables
\begin{eqnarray}
\alpha_t & = & \frac 1 2 [  \frac 1 2 - {\cal H}  ] , \\
\alpha_{\theta} & = & - \frac 1 2 X^{A'} \Pi_A.
\end{eqnarray}
As a consequence of the form of the Hamiltonian the function
$\alpha_\theta$ satisfies the identity
\begin{equation}
\frac{\partial \alpha_\theta}{\partial X^{A'}}
\left( \frac{\delta H}{\delta \Pi_{A}} \right)'
- \frac{\partial \alpha_\theta}{\partial \Pi_{A}}
\frac{\delta H}{\delta X^{A}} =-\frac{1}{2}
{\cal H}'=\alpha_{t,\theta}   \label{derivativealpha}
\end{equation}
for arbitrary functions $X^A$ and $\Pi_A$. For $X^A$ and $\Pi_A$ that
satisfy the Hamilton's equations of motion the left-hand side of
\reff{derivativealpha} becomes the time derivative of $\alpha_\theta$
and we obtain
\begin{equation}
 \alpha_{\theta},_t =\alpha_t,_{\theta} \quad {\rm or} \quad
d\alpha=0,
\label{grad}
\end{equation}
which guarantees that the line integral does not depend on the path.
To have the function $A$ globally defined on the $t\theta$ cylinder we
must impose one more constraint on $W$ and $\Phi$, namely that $A$
must also be periodic in $\theta$ with period 2$\pi$, which leads to
the constraint
\begin{equation}
\int_0^{2\pi} \alpha_{\theta}(t,\theta)  d\theta = 0 .
\end{equation}
This constraint has to be imposed only at one time $t_0$ because it is
conserved during the time evolution due to equation \reff{grad}.

Two important subclasses of Gowdy spacetimes are obtained by imposing
special conditions that are conserved by the evolution equations. We
shall refer to the spacetimes satisfying the condition $\Phi=0$ as the
``polarized'' Gowdy spacetimes and to the spacetimes satisfying \mbox{
$ W=W(t)$ and $\Phi=n \theta$}, for some $n\in Z$, as the ``circular
loop'' spacetimes. For the polarized case the evolution equations for
$W$ and $\Phi$ reduce to one {\em linear} partial differential
equation, and for the circular loop case to  one non-linear {\em
ordinary} differential equation.

To facilitate further analysis we will introduce new coordinates $x$
and $y$ in the $W\Phi$ hyperbolic space, such that
\begin{eqnarray}
\tanh W & = & 2 \sqrt{ \frac{x^2 + ( y-1)^2}{ x^2 + (y+1)^2} }  \\
\tan \Phi & = & \frac{x^2 + y^2 -1}{2 x}.
\end{eqnarray}
These are just the Poincar\'e half-plane coordinates in which the
hyperbolic metric becomes
\begin{equation}
g = \frac{dx^2 + dy^2}{y^2}, \label{half-plane}
    \; x \in R, \; y > 0.
\end{equation}
The equations of motion for $x$ and $y$ are also given by Hamilton's
equations where now, however, the metric in \reff{hamiltonian} is
replaced by expression \reff{half-plane} and $\{X^1,X^2\}$ now stand
for $\{x,y\}$. With this modification the Hamiltonian density becomes
a rational function of its arguments. The functions appearing in the
Gowdy metric \reff{gowdymetric}, expressed in terms of the $x$ and
$y$, are
\begin{eqnarray}
\cosh W + \cos \Phi \sinh W = \frac{(x+y)^2 +1}{2y} \nonumber\\
\cosh W - \cos \Phi \sinh W = \frac{(x-y)^2 +1}{2y}
\label{metricfun}\\
\sin \Phi \sinh W = \frac{x^2+y^2 -1}{2y} \nonumber.
\end{eqnarray}

Note that the $y>0$ restriction is necessary to obtain a metric
\reff{gowdymetric} that is of Lorentzian signature and that the
transformation $x\rightarrow -x$ gives an isometric metric ( it
amounts to exchange of $x^1$ and $x^2$).

\section{The Perturbative Expansion}

\subsection{The Perturbative Method}
As already mentioned, in general we do not know how to solve the
evolution equations for $x$ and $y$.  What we can do, however, is try
to solve the equations perturbatively to all orders and analyse the
asymptotic behavior of the expansion to learn about the behavior of
the Gowdy space-times close to the singularity at $t=\infty$.  To
facilitate this analysis  we will first introduce a seemingly even
more complicated problem. We will try to solve the evolution equations
for the $\epsilon$ dependent Hamiltonian density
\begin{equation}
{\cal H}_{\epsilon}   =  \frac{1}{2} [ g^{AB} \Pi_A \Pi_B +
                                  \epsilon \; e^{-2t} g_{AB} X^{A'}
                             X^{B'}  ] . \label{ehamiltonian}
\end{equation}
The case of real interest to us is $\epsilon=1$,  but it is easy to
see  that any solution of the $\epsilon$-dependent equations for some
$\epsilon_1 > 0$ is also a solution of the equations for any other
$\epsilon_2 > 0$ , provided we shift the time $t$ by an appropriate
constant. Precisely stated, if $X_{\epsilon_1}(t)$ is a solution of
the equations of motion for \mbox{$\epsilon=\epsilon_1 >0$} then
\begin{equation}
X_{\epsilon_2}(t) = X_{\epsilon_1}(t- \frac{1}{2}
\ln \epsilon_2 / \epsilon_1)
\end{equation}
is a solution of the equations for $\epsilon=\epsilon_2>0$. So, with
the introduction of $\epsilon$ we have not changed any significant
property of the problem, but the $\epsilon$ in the Hamiltonian will
help us obtain simple recursive form for  the perturbative
calculations.

Assume that there exists an $\epsilon$--dependent family
$X(\epsilon,t,\theta)$  of solutions to the modified equations which
depends smoothly upon the parameter , in particular around
$\epsilon=0$, and  expand it into a power series in $\epsilon$ to find
evolution equations for each term in the expansion:
\begin{eqnarray}
X(\epsilon,t,\theta)& = & \sum_{n=0}^{\infty} \frac{\epsilon^n}{n!}
                         X^{(n)}(t,\theta) \label{expansion}\\
X^{(n)}(t,\theta) & = & \left. \frac{d^n}{d\epsilon^n}
               X(\epsilon,t,\theta) \right|_{\epsilon=0} \\
                X & = &
\left( \begin{array}{c} x \\ y \\ \dot x \\ \dot y \\ \end{array}
\right)
\end{eqnarray}
Whether the family actually exists depends on the convergence
properties of the ``formal'' perturbative expansion \reff{expansion}
which can be changed using the freedom in choosing appropriate
$X^{(n)}$ in every order since the $n$-th order perturbation equations
don't have unique solutions; but as we shall see,irrespective of the
convergence of the expansion, the recursive calculations of the higher
order terms are always well defined in terms of the lower order terms.
It could easily happen that the expansion could not be made convergent
but could nevertheless be made asymptotic as $t\rightarrow \infty$, in
which case we would still be able to study the behavior of Gowdy
spacetime close to the singularity. The question still remains whether
we can study all, or almost all, Gowdy spacetimes using this
perturbative method; i.e. can we for any Gowdy solution find an
$\epsilon$-dependent family which for $\epsilon=1$ reduces to that
particular Gowdy solution and is smooth around $\epsilon=0$? We will
return to this question later, for now let us just say that a function
counting argument, together with the properties of our expansion,
suggests that the expansion is possible for an open subset of %may be
too strong all Gowdy  solutions on $T^3 \times R$.

To clarify the analysis let us  for a moment consider an s-dimensional
system of nonlinear partial differential equations on $R \times M$,
with ``time'' coordinate in $R$ and ``space'' coordinates in an
arbitrary finite dimensional manifold $M$. In shorthand notation we
write
\begin{equation}
\dot X = h(\epsilon,X,Y)=
g(X) + \epsilon \; f(Y) , \label{generalform}
\end{equation}
where $X$ is an s-component function and $Y$ is a multicomponent
function that contains all the components of $X$ and some number of
partial derivatives of $X$ with respect to the ``spatial''
coordinates. For $\epsilon=0$ this system reduces to a system of
ordinary differential equations. By differentiating \reff{generalform}
with respect to $\epsilon$ and then putting $\epsilon=0$ we obtain the
equations that have to be satisfied by $X^{(n)}$ if a family
$X(\epsilon)$ exists:
\begin{eqnarray}
\dot X^{(n)}& =& \left. \frac{d^n}{d\epsilon^n} h(\epsilon,X,Y)
\right|_{\epsilon=0}\equiv h^{(n)}
 \Longrightarrow \label{perturbative:eq} \\
 n &=&0 \qquad \dot X^{(0)} = g(X^{(0)})  \\
n&>& 0 \qquad
\dot X^{(n)} = g^{(n)} + n f^{(n-1)}
\end{eqnarray}
Now we will analyse the above perturbative equations without worrying
about the existence of an $\epsilon$-dependent family of solutions of
the equations \reff{generalform}. The derivatives of $g(X)$ and $f(Y)$
can be calculated using the chain rule, and  with all the indices
explicitly written, the expression for $g(X)$ is
\begin{eqnarray}
g_j^{(n)}&\equiv&
\left. \frac{d^n}{d\epsilon^n} g_j(X) \right|_{\epsilon=0}  =
\nonumber \\
&=&\sum_{k=1}^{n} \sum_{i_1 \cdots i_k}
                   \partial^k g_{j i_1 \cdots i_k}(X^{(0)})
\left[    \sum_{
 \scriptstyle  m_1 \cdots m_k \atop \scriptstyle \sum m_k = n
               }
X_{i_1}^{(m_1)} \cdots X_{i_k}^{(m_k)} \right]   \label{gn} \\
&& \partial^k g_{j i_1 \cdots i_k}(X^{(0)}) \equiv
\left. \frac{\partial^k g_j(X)}{\partial X_{i_1} \cdots
                              \partial X_{i_k} } \right|_{\epsilon=0}.
\end{eqnarray}
{}From now on we will suppress the vector indices, like $ji_1 \cdots
i_k$, and arguments of $\partial^n g \;$ and $\partial^n f \;$ to
simplify the notation. For the same reason we will put $g(X^{(0)})
\equiv g^{(0)}$ and  $f(Y^{(0)}) \equiv f^{(0)}$. In this shorthand
notation the evolution  equations for $X^{(n)}$ for any $n \geq 1$ are
\begin{eqnarray}
\dot X^{(n)}  =  \partial g \; X^{(n)} + \sum_{k=2}^n \partial^k g \;
\left[    \sum_{
 \scriptstyle  m_1 \cdots m_k \atop \scriptstyle \sum m_k = n
               }
X^{(m_1)} \cdots X^{(m_k)} \right]  \nonumber  \\
+  \delta_{1n} f^{(0)} + n \sum_{k=1}^{n-1} \partial^k f \;
\left[    \sum_{
 \scriptstyle  m_1 \cdots m_k \atop \scriptstyle \sum m_k = n-1
               }
Y^{(m_1)} \cdots Y^{(m_k)} \right]
\end{eqnarray}
The first three equations  including the zeroth are
\begin{eqnarray}
n=0 \quad & \dot X^{(0)}  = & g(X^{(0)}) \label{order0} \\
n=1 \quad & \dot X^{(1)}  = & \partial g \; X^{(1)} + f(Y^{(0)})
\label{order1} \\
n=2 \quad & \dot X^{(2)}  = & \partial g \; X^{(2)} + \partial^2 g \;
     X^{(1)} X^{(1)} +  2  \partial f \; Y^{(1)}
\end{eqnarray}

We see that to  all  orders, except the zeroth, the perturbative
equations are ordinary inhomogeneous linear differential equations of
the form \begin{equation} \dot X^{(n)} =\partial g \; X^{(n)} +
S^{(n)}, \label{ordern} \end{equation} where the matrix that
multiplies $X^{(n)}$ is the same for all orders and the source
$S^{(n)}$ depends only on the lower order functions $X^{(k)}$ and
$Y^{(l)}$, i.e. $k,l<n$. Once we know the general solution for the
zeroth order equation \reff{order0}, which we shall refer to as  the
seed solution, we can recursively solve all the higher order equations
using the ``variation of constants'' method for ordinary linear
differential equations. The general solution of the $n$-th order
equation
\reff{ordern} is
\begin{equation}
X^{(n)}(t) = X_{hom}^{(n)}(t) +
 \int_{t_0}^t {\cal G}(t,t') \;  S^{(n)}(t') dt', \label{int}
\end{equation}
where $X_{hom}^{(n)}$ is a solution of the homogeneous part of
Eq.\reff{ordern} and $\cal G$ is its Green's function. Both can be
calculated from the seed solution. By differentiating the seed
solution with respect to the constants $c_i$ it depends on, and
linearly combining these derivatives, we obtain the general
homogeneous solution
\begin{equation}
X_{hom} = \sum_{i=1}^s \alpha_i \frac{\partial X^{(0)}}{\partial c_i},
\end{equation}
where $\alpha_i$ are arbitrary constants.
The Green's function ${\cal G}(t,t')$ for \reff{int}
is the Jacobian matrix for the transformation
\begin{equation}
\frac{\partial  X^{(0)}(t)}{\partial X^{(0)}(t') } \label{green}
\end{equation}
and can be calculated from the seed solution expressed in terms of
the initial values.
It is easy to check that
\begin{equation}
\dot X_{hom} = \partial g \; X_{hom}
\end{equation}
and
\begin{equation}
\frac{\partial}{\partial t} {\cal G}(t,t') =
\partial g(t) \; {\cal G}(t,t')
\qquad \forall t',
\end{equation}
and confirm that \reff{int} satisfies \reff{ordern}. The lower limit
of integration in \reff{int} could have been chosen to be different
at every order as well, but a change in its value corresponds just to
adding a solution of the homogeneous equation.

Using any finite number of the solutions $X^{(n)}$ of the perturbative
equations \reff{perturbative:eq} we can construct an
$\epsilon$-dependent function
\begin{equation}
X_N(\epsilon) = \sum_{n=0}^N \frac{\epsilon^n}{n!} X^{(n)},
\end{equation}
which we shall refer to as the formal expansion of $X$ .
For any formal expansion we can define a function
\begin{equation}
\delta_N(\epsilon) = \dot X_N(\epsilon) - h(\epsilon,X_N),
\end{equation}
which, as a consequence of perturbative equations, satisfies
\begin{equation}
\frac{d^n}{d\epsilon^n}
\left. \delta_N(\epsilon)\right|_{\epsilon=0} \equiv
\delta_N^{(n)} = 0, \qquad \forall n\leq N. \label{delta}
\end{equation}
To simplify the notation we have assumed that $h$ depends only on $X$.
Given a formal expansion $X_N$ of $X$, we can generate a formal
expansion $F_N$ of any smooth function $F$ by
\begin{equation}
F_N(\epsilon) = \sum_{n=0}^N \frac{\epsilon^n}{n!} F^{(n)}, \qquad
F^{(n)}=
\left. \frac{d^n}{d \epsilon^n}  F(X_N(\epsilon))\right|_{\epsilon=0},
\label{Fexpansion}
\end{equation}
where $F^{(n)}$ in general depends on all $X^{(k)}$, with $k\leq n$,
and is explicitly given by \reff{gn} with $g$ replaced by $F$.

An interesting consequence of the above definition and Eq.\reff{delta}
is that if we have a constant of motion for the evolution governed by
\reff{generalform}, i.e. a function $C(X)$ such that identically for
any $X$ and $\epsilon$
\begin{equation}
\frac{\partial C(X)}{\partial X}\; h(\epsilon,X)=0, \label{cmotion}
\end{equation}
all the terms $C^{(n)}$ in the formal expansion of $C$ will
be constant in time
also. To see that, consider a formal expansion $C_N$ of $C$.
Using \reff{cmotion} we can evaluate
\begin{eqnarray}
\dot C(X_N) &= &\frac{\partial C(X_N)}{\partial X} \dot X_N
= \frac{\partial C(X_N)}{\partial X}
\left[ h(\epsilon,X_N) + \delta_N(\epsilon) \right]  \nonumber \\
&=& \frac{\partial C(X_N)}{\partial X} \delta_N(\epsilon),
\end{eqnarray}
differentiating $n$ times with respect to $\epsilon$ and using
\reff{delta} we obtain
\begin{equation}
\dot C^{(n)} = \left. \frac{d^n}{d \epsilon^n}
\dot C(X_N(\epsilon))\right|_{\epsilon=0} =0.
\label{allconstant}
\end{equation}
The general idea of using $\delta_N^{(n)}=0$ could be employed to
find other properties of the perturbative equations that are inherited
from the ``full'' equations.

To conclude this general discussion we once more note that one has
considerable freedom in choosing the formal expansion apart from the
freedom in the choice of the seed solution. One can use this freedom
to try to obtain an expansion in which, when time goes to some $t_0$,
all  terms in the expansion form an asymptotic sequence of functions
i.e.
\begin{equation}
\frac{X^{(n+1)}(t)}{X^{(n)}(t)} \rightarrow 0 \qquad t\rightarrow t_0,
\label{asymptotic}
\end{equation}
for all $n$; or perhaps even to obtain a convergent expansion.

Since we are interested in finding the asymptotic behavior of
the solutions of the Gowdy equations
when $t\rightarrow \infty$, we shall chose the lower limit of
integration in \reff{int} to be $t_0=\infty$
and let $X_{hom}^{(n)}=0$ for all
$n$. This choice leads uniquely to the fastest possible
decay of the higher order
terms, because the homogeneous solutions of the perturbative Gowdy
equations do not decay faster than the seed solution when
$t\rightarrow
\infty$.
The $X^{(n)}$ in \reff{int} after shifting the integration variable
$t'\rightarrow t'-t$  becomes
\begin{equation}
X^{(n)}(t) = -\int_{0}^{\infty} {\cal G}(t,t+t') \;
S^{(n)}(t+t') dt',
                \label{int0inf}
\end{equation}
and depends only on the seed solution. As we shall see, this choice
leads to a uniformly behaved asymptotic sequence for a large range of
seed solutions, which suggests that the sequence  might be an
asymptotic expansion of an exact Gowdy solution.

\subsection{The Geodesic Loop Solution and the First Order Correction}
Returning to the Gowdy equations we note that Hamilton's equations for
the Hamiltonian \reff{ehamiltonian} are of the form \reff{generalform}
with
\begin{eqnarray}
X= \left( \begin{array}{c} x \\ y\\ \dot x \\ \dot y \end{array}
   \right)
& &Y= \left( \begin{array}{c} x \\ y\\  x' \\  y' \\ x'' \\ y''
\end{array}
  \right)
\end{eqnarray}
and
\begin{eqnarray}
g(X) = \left( \begin{array}{c} X_3 \\ X_4 \\ A \\ C - B \end{array}
   \right)
& & f(Y) =  e^{-2t}
   \left( \begin{array}{c} 0 \\ 0 \\ D - E \\ F+G-H \end{array}
   \right) \label{fandg}
\end{eqnarray}
where
\begin{equation}
\begin{array}[t]{l}
\displaystyle
\dis A= \frac{2}{X_2} X_3 X_4  \\
\dis B= \frac{1}{X_2} X_3^2   \\
\dis C= \frac {1}{X_2} X_4^2
\end{array}
\qquad
\begin{array}[t]{l}
D= Y_5  \\
\dis E= \frac {2}{Y_2} Y_3 Y_4  \\
F= Y_6  \\
\dis G= \frac {1}{Y_2} Y_3^2   \\
\dis H= \frac {1}{Y_2} Y_4^2
\end{array}
\end{equation}

The zeroth order equations are, as is obvious from the
Hamiltonian~\reff{ehamiltonian}, just the geodesic equations in
the Poincar\'e
half--plane whose  general solution, using  our notation, is
\begin{eqnarray}
 X_1^{(0)}& =& a + b \tanh(ct+d) \label{gloop}\\
 X_2^{(0)}& =& b \frac{1}{\cosh(ct+d)} \\
 X_3^{(0)}& =& bc \frac{1}{\cosh^2(ct+d)} \\
 X_4^{(0)}& =& - bc \frac{\sinh(ct+d)}{\cosh^2(ct+d)},
\end{eqnarray}
where $a$, $b>0$, $c> 0$ and $d$ are time independent but otherwise
arbitrary smooth functions of $\theta$. For convenience we shall refer
to these as the ``geodesic loop solutions'', since each curve
$X^{(0)}(t,\theta_0)$, for fixed $\theta_0$, is a geodesic. The speed
of these geodesic loops, evaluated using Eq. \reff{speed}, is equal to
the function $c(\theta)$ and is, as it should be, constant in time.
The geodesic loop solutions have the same number of free initial data
as the general Gowdy solutions. Note that when $a,b,c$ and $d$ do not
depend on $\theta$, the geodesic loop, in this case degenerated to
just one point, is a solution to the full Gowdy equations. The
geodesic loop solutions give rise to the ``geodesic loop spacetimes''
which, we hope to prove, are asymptotically (as $t\rightarrow \infty$)
approached by Gowdy spacetimes.

Differentiating the geodesic loop solutions with respect to $a,b,c$
and $d$ we obtain the general homogeneous solution for the
perturbative Gowdy equations. It is easy to see that it does not decay
faster than the seed solution when $t \rightarrow\infty$, which
justifies  our choice of $X^{(n)}$ in \reff{int0inf} as unique for a
(potentially) well behaved asymptotic expansion.

Since we are interested in the asymptotic behavior when $t
\rightarrow\infty$ we will expand the above functions in power series
in $1/\xi$, where $\xi = e^{(ct+d)}$, which are convergent for all
$\xi >1$ i.e. for all $t$ and $\theta$ for which $ct+d>0$. To exhibit
the asymptotic behavior we will factor out the lowest power of $\xi$.
For example:
\begin{eqnarray}
X_1^{(0)}&=&a + b + \sum_{k=1}^{\infty} 2b (-1)^k  \xi^{-2k} \\
X_2^{(0)}& =& \xi^{-1}  \sum_{k=0}^{\infty} 2b (-1)^k  \xi^{-2k}.
\end{eqnarray}
In fact all the the zeroth order functions
have a similar form. The other two components of $X^{(0)}$ are  time
derivatives of the first two, and $Y^{(0)}$ includes also $\theta$
derivatives of $X^{(0)}$.
\begin{definition}
Let F:R $\rightarrow$ R be a function that can be expanded into a
uniformly convergent series
\begin{equation}
F(t) = \xi^{-N} \sum_{k=0}^{\infty} a_k \xi^{-2k}, \label{xiexpansion}
\end{equation}
where $\xi =e^{(ct+d)}$ , $N \in R$ , all $a_k$ are polynomials in $t$
of degree less than or equal to some fixed integer $p\geq0$ and
$a_0\neq0$. Then, for easy reference, we call $F$ a
\mbox{$\xi$--expandable} function , $N(F)$ the decay exponent of $F$,
p(F) the polynomial degree of F and $a_0(F)$ the dominant coefficient
of $F$. If $F(t)=0$ for all $t$ we put $N(F)=\infty$. \end{definition}
This definition will be important in the  inductive proof of the decay
of all higher order terms since all the functions we will be dealing
with in this paper satisfy the conditions of the above definition,
i.e. they are \mbox{$\xi$--expandable}. For any two such functions $F$
and $G$ the product $FG$ is obviously a $\xi$--expandable function,
with $p(FG)=p(F)+p(G)$ and with $N(FG) = N(F) + N(G)$. Moreover, if
$N(F)-N(G)=2k$ for some integer $k>0$ the sum $F+G$ will also be
$\xi$--expandable and $N(F+G)=N(G)$. In the special case when
$N(F)=N(G)$ we have $N(F+G)\geq N(F)$, because of the possible
cancellation of the dominant coefficients. If $k$ dominant
coefficients cancel we have $N(F+G)= N(F)+ 2 k$. So, in general, for
any two $\xi$-expandable functions $F$ and $G$ whose sum is also
$\xi$-expandable, we have
\begin{equation}
N(F+G) \geq {\rm min}\{N(F),N(G)\} \label{sumineq}
\end{equation}

Owing to the uniform convergence of expansion \reff{xiexpansion}
the integral
\begin{equation}
\int_0^{\infty} F(t) dt = \int_0^{\infty}
         \sum_{k=0}^{\infty} a_k \xi^{-2k-N} dt
=   \sum_{k=0}^{\infty} \int_0^{\infty} a_k \xi^{-2k-N} dt,
\label{tbyt}
\end{equation}
for all $F$ for which $N(F)\geq0$. Therefore, to evaluate
\reff{int0inf} we need to integrate
only simple integrals of the form
\begin{equation}
\int_0^{\infty} t^r e^{-(2k+N)ct} dt =
\frac{r!}{[(2k+N)c]^{r+1}},
\end{equation}
where the integer $k\geq0$ and the integer $r\leq p(F)$, the
polynomial degree of $F$. Note that all the zeroth order functions
$X^{(0)}$ have $p(X^{(0)})=0$, and that each differentiation with
respect to $\theta$ brings one $t$ down from the exponent and
therefore increases the polynomial degree by 1.

In the generic case, i.e. when no special conditions are imposed on
the functions $a$, $b$, $c$ and $d$, the decay exponents for the
zeroth order functions are
\begin{equation}
N(X^{(0)}) = \left(
\begin{array}{c} 0 \\ 1 \\ 2 \\ 1 \end{array}
             \right)
\qquad
N(Y^{(0)}) = \left(
\begin{array}{c} 0 \\ 1 \\ 0 \\ 1 \\ 0 \\ 1 \end{array}
             \right),
\end{equation}
from which follows that
\begin{equation}
\begin{array}{rcl}
N(A)=2& &N(D)=0 \\
N(B)=3& &N(E)=0 \\
N(C)=1& &N(F)=1 \\
{}    & &N(G)=-1 \\
{}    & &N(H)=1.
\end{array}
\end{equation}

{}From \reff{order1} and \reff{fandg} we can calculate the first order
source vector $S^{(1)}(t)$; its decay exponents are
\begin{equation}
N(S^{(1)})= N(f^{(0)})= \left(
\begin{array}{l} \infty \\ \infty\\ \lambda+2 \\ \lambda+1 \end{array}
                        \right),\label{exponentS1}
\end{equation}
where $\lambda = 2(1-c)/c$, and its polynomial degree $p(S^{(1)})=2$.

{}From the form of Gowdy's $g$ and $f$ we see that the first two
components of the 4-dimensional source vector $S^{(k)}$ are zero for
all $k$, which implies that we need just the last two columns of the
Green's function to evaluate integrals \reff{int0inf}. Explicit
calculation of the Green's function $4\times4$ matrix ${\cal
G}(t,t+t')$ shows that it depends on $\theta$ only through the
functions $c$ and $d$, and that all its components are of the form
\reff{xiexpansion} with the coefficients $a_k$ polynomials of the
first order in $t'$ multiplied by $e^{nt'}$, with $n$ ranging from -1
to 2. The time $t$ appears only in the exponents, therefore the
polynomial degree of $\cal G$ is zero. Asymptotically for
$t\rightarrow \infty$ we obtain
\begin{equation}
{\cal G} \sim \left(
\begin{array}{cccc}
\; *&*& (1-\zeta^2)/2c & \xi^{-1}
(\zeta^{-1} -\zeta + 2\zeta  c t')/c \\
\; *&*& \xi^{-1} (1-\zeta^{2} +2ct')/c & -\zeta t' \\
\; *&*& \zeta^{2}        & -\xi^{-1} 4\zeta c t' \\
\; *&*& \xi^{-1} (3\zeta^{2} - 2ct' - 3) & \zeta (ct'+1)
\end{array}
              \right).
\end{equation}
Here we have put asterisks for uninteresting components and we have
written $\zeta$ for $e^{ct'}$. The highest power of $\zeta$ in the
components of the Green's function will be important for the
convergence properties of the integral \reff{int0inf}, and it is
convenient to introduce another simple definition to keep track of the
exponential functions of the integration variable $t'$. Let $N'(F)$
denote the function defined like $N(F)$ in Definition 1, with $t$
changed to $t'$ and with $\zeta=e^{ct'}$ written in place of
$\xi=e^{(ct+d)}$. All the functions we use obviously satisfy the
conditions stated in Definition 1, with respect to both $t$ and $t'$.
Pairing the exponents $N$ and $N'$ in parentheses for brevity, from
the exact expression for the Green's function we obtain
\begin{equation}
(N,N')({\cal G}) = \left(
\begin{array}{cccc}
\; *&*& (0,-2) & (1,-1) \\
\; *&*& (1,-2) & (0,-1) \\
\; *&*& (0,-2) & (1,-1) \\
\; *&*& (1,-2) & (0,-1)
\end{array}
              \right).
\end{equation}
Because the identity $N(F(t+t'))=N'(F(t+t'))$ is valid for any
function satisfying the conditions in Definition 1, using
\reff{exponentS1}, we obtain
\begin{equation}
(N,N')(S^{(1)}) = \left(
\begin{array}{l} (\infty,\infty) \\ (\infty,\infty) \\
 (\lambda+2,\lambda+2) \\ (\lambda+1,\lambda+1) \end{array}
                        \right).
\end{equation}
Multiplying the Green's function and the source we find the integrand
whose exponents are
\begin{equation}
(N,N')({\cal G} S^{(1)}) =\left(
\begin{array}{l} (\lambda+2,\lambda) \\ (\lambda+1,\lambda) \\
 (\lambda+2,\lambda) \\ (\lambda+1,\lambda) \end{array}
                        \right).
\end{equation}
The integral will converge for all $\theta$ for which
\begin{equation}
\lambda(\theta) > 0    \Longleftrightarrow c(\theta) \in (0,1).
\end{equation}
Finally, the integration gives the first order functions which
are, because of \reff{tbyt}, also $\xi$--expandable. We were
able to calculate the first order integral explicitly, without
restoring to the $\xi$-expansion, and thus obtain the exact
expressions for the first order functions, whose asymptotic form
agreed with the result obtained by $\xi$-expansion. The polynomial
degree of $X^{(1)}$ is 2, the same as $p(S^{(1)})$, since
$p({\cal G})=0$, but the dominant coefficient of all $X^{(1)}$-s are
just first degree polynomials in $t$.
The differences of the decay exponents of the first  and zeroth order
functions are
\begin{equation}
N(X^{(1)})-N(X^{(0)}) = \left(
\begin{array}{c} \lambda+2 \\\lambda \\ \lambda \\ \lambda \end{array}
             \right),
\quad
N(Y^{(1)})-N(Y^{(0)}) = \left(
\begin{array}{c} \lambda+2 \\ \lambda \\ \lambda+2 \\ \lambda \\
\lambda+2 \\ \lambda \end{array}
             \right), \label{decay1}
\end{equation}
and, again, $\lambda>0$ is the condition that has to be satisfied if
we want that
\begin{equation}
\frac{X_i^{(1)} }{X_i^{(0)}} \sim t \xi^{-\lambda} = t e^{-2 (1-c)t}
\rightarrow 0 \quad {\rm when}\;
 t \rightarrow \infty,
\end{equation}
for $i=1,2,3$, ( $X_1^{(1)}$ decays even faster).

The condition $c(\theta)<1$ can be relaxed for $\theta$ in some  $I_0
\subset [0,2 \pi)$, if $a(\theta)+b(\theta)=$const. for $\theta \in
I_0$. Explicit calculation in this case shows that the decay exponents
of the first order source increase by 2 and the first order integral
converges for all $c(\theta)>0$ , $\theta \in I_0$ giving
\begin{equation}
N(X^{(1)})-N(X^{(0)}) = \left(
\begin{array}{c} \lambda+4 \\\lambda+2 \\ \lambda+2 \\ \lambda+2
\end{array}
             \right),
\quad
N(Y^{(1)})-N(Y^{(0)}) = \left(
\begin{array}{c} \lambda+4 \\ \lambda+2 \\ \lambda+4 \\ \lambda+2 \\
\lambda+4 \\ \lambda+2 \end{array}
             \right),
\end{equation}

Now when we know the decay exponents of the zeroth and first order
functions we can go on to calculate the decay exponents of the higher
order functions using \reff{int0inf}, but all we actually need to know
is whether they increase with order.

\subsection{Higher Order Terms} The lower bounds for the decay
exponents of $X^{(m)}$ can be easily found for any order $m$ following
a rather simple inductive argument. \begin{theorem} Let $X^{(m)}$ be a
solution of the perturbative Gowdy equations given by \reff{int0inf},
and let its geodesic loop parameters a, b, c and d be arbitrary smooth
functions of $\theta$ except that $c(\theta) \in (0,1)$, then for any
$m\geq1$, the decay exponent of $X^{(m)}$ satisfies the inequality
\begin{equation}
N(X^{(m)}) \geq N(X^{(1)}) + (m-1) \lambda. \label{theorem}
\end{equation}
\end{theorem}
{\bf Proof:} \quad The statement of the theorem is trivially true for
$m=1$.

Inductive hypothesis: Assume
that the statement \reff{theorem} is true for all $1\leq m \leq n-1$.
Using the result \reff{decay1} from the previous section we obtain
\begin{equation}
N(X^{(m)}) \geq N(X^{(0)}) + m \lambda, \label{decaym}
\end{equation}
for all  $1\leq m \leq n-1$.
Now, all we need to prove the theorem is show that the
decay exponents of the $n$-th order source $S^{(n)}$ satisfy the
inequality
\begin{equation}
N( S^{(n)} ) \geq  N( S^{(1)} ) + (n-1) \lambda \label{sourceineq}
\end{equation}
which will, because of the inequality \reff{sumineq},
after multiplication with the Green's function and integration
over $t'$ give the desired result \reff{theorem} for $m=n$.

The components of the source vector $S^{(n)}$ for $n>1$ are sums of
parts of the form
\begin{eqnarray}
S^{(n)}(g)& +& S^{(n)}(f) \\
S^{(n)}(g)&=& \left. \partial^k g \; X^{(m_1)} \cdots X^{(m_k)}
\right|_{\sum m_i =n}
\\
S^{(n)}(f)&=& \left. \partial^k f \; Y^{(m_1)} \cdots Y^{(m_k)}
\right|_{\sum m_i =n-1}
\label{sourceterm}
\end{eqnarray}
where for $g$ we can substitute $A$, $B$ or $C$ and for $f$ $D$, $E$,
$F$, $G$ and $H$. Again, because of the inequality \reff{sumineq}, we
need to find only  the minimal decay exponent  of any term in the sum
to obtain a lower bound to the decay exponent of $S^{(n)}$. To do this
we need to take a closer look at the form of the terms above.

All the functions $A,B,C,D,E,F,G$ and $H$  are of the form $P=\Pi_j
Z_j^{k_j}$, where $Z$ stands for $X$ and $Y$, e.g. $A=2 X_2^{-1} X_3
X_4$. Derivatives of such functions are easy to evaluate and we have
\begin {equation}
\frac{\partial P}{\partial X_j} =
\frac{k_j}{X_j} P \Rightarrow N(\frac{\partial P}{\partial X_j}) =
N(P) - N(X_j)
\end{equation}
if $k_j\neq0$, otherwise $N=\infty$.
Using the above property we find
\begin{eqnarray}
N(\partial^k f \; Y^{(m_1)} \cdots Y^{(m_k)})& =& N(f^{(0)})
\nonumber \\
& + &
\sum_{\scriptstyle i=1\atop \sum m_i=n-1 \scriptstyle }^k
 [N(Y^{(m_i)}) - N(Y^{(0)})] \\
N(\partial^k g \; X^{(m_1)} \cdots X^{(m_k)})& =& N(g^{(0)})
\nonumber \\
& +&
\sum_{\scriptstyle i=1\atop \sum m_i=n \scriptstyle  }^k
 [N(X^{(m_i)}) - N(X^{(0)})]
\end{eqnarray}
In the sums above only $X$-s and $Y$-s of the order  $n-1$ or lower
appear and the inequality \reff{decaym} together with the
conditions  $\sum m_i=n-1$ and  $\sum m_i=n$, for $Y$ and $X$ sums
respectively, enable us to conclude that
the sums satisfy the following inequalities
\begin{eqnarray}
\sum_{i=1}^k [N(Y^{(m_i)}) - N(Y^{(0)})] &\geq& (n-1) \lambda,
\label{fsum} \\
\sum_{i=1}^k [N(X^{(m_i)}) - N(X^{(0)})] &\geq& n \lambda.
\label{gsum}
\end{eqnarray}
Recalling that $S^{(1)}=f^{(0)}$ we obtain
\begin{equation}
N(S^{(n)}(f)) \geq N(S^{(1)})+(n-1)\lambda,
\end{equation}
and using $N(g^{(0)})=N(f^{(0)}) - \lambda$ and \reff{gsum} we get
\begin{equation}
N(S^{(n)}(g)) \geq N(S^{(1)})+(n-1)\lambda.
\end{equation}
Therefore, the source $S^{(n)}$ has the desired behavior, which
concludes the proof.

When no cancellations in the dominant coefficients are present in the
course of calculations, the statement of the theorem becomes an
equality, and for $n\geq 1$ and for $t \rightarrow \infty$ we obtain
\begin{eqnarray}
x^{(n)}&=&X_1^{(n)} \sim p_{2n} e^{-2n(1-c)t} e^{-2t} \\
y^{(n)}&=&X_2^{(n)} \sim q_{2n} e^{-2n(1-c)t} e^{-ct}.
\end{eqnarray}
The $p_{2n}$ and $q_{2n}$ are polynomials in $t$ of degree $2n$ or
possibly lower, which can be easily shown using an argument similar to
the one used for the decay exponents. We see that the terms decay
faster and faster with order and that we have obtained an asymptotic
sequence. When cancellations are present, the decay is even faster and
we can group the terms to obtain an asymptotic sequence again.

A statement similar to the Theorem 1. is true for the case
$a+b$=const. and $c$ arbitrarily large. Then the higher order terms
decay even faster, and their decay exponents satisfy the inequality
\begin{equation}
N(X^{(n)}) \geq N(X^{(1)}) + (n-1) (\lambda+2)
\end{equation}

\subsection{Expansion of the Metric Functions and Curvature}

Having obtained the asymptotic sequence of solutions to the
perturbative equations for the Gowdy functions $x$ and $y$ we can
generate the expansion for any smooth function of $x$ and $y$ using
\reff{Fexpansion}. For example, let $F$ be a function of $X$ of the
form
\begin{equation}
F(X)=\frac{P(X)}{Q(X)},
\end{equation}
where $P$ and $Q$ are polynomial functions of $X$.
To find the expansion
\begin{equation}
F_M=F(X^{(0)})+ \sum_{n=1}^M \frac{\epsilon^n}{n!} F^{(n)}
\end{equation}
to arbitrary finite order $M$, we have to calculate the
$F^{(n)}$-s. They
are given in terms of $\partial^k F$ by the equation \reff{gn},
with $g$ replaced by $F$. The first partial derivative of $F$,
\begin{equation}
\partial F = \frac{P}{Q} \left[
 \frac{\partial P}{P} -  \frac{\partial Q}{Q}
\right], \label{F}
\end{equation}
is again a function of the form $\tilde P/ \tilde Q$, with $\tilde P$
and $\tilde Q$ polynomials in $X$. For each monomial $T$ in, say, $P$
we have \mbox{$N(\partial T) = N(T)-N(X)$}, which using \reff{sumineq}
gives $N(\partial P)\geq N(P)-N(X)$. Since \mbox{$N(\partial P /
P)=$}\mbox{$N(\partial P)-N(P)\geq -N(X)$}, using \reff{F} we obtain
\begin{equation}
N(\partial F) \geq N(F)-N(X),
\end{equation}
and repeating the same reasoning $k$ times,
\begin{equation}
N(\partial^k F) \geq N(F)-k N(X). \label{NFk}
\end{equation}
Using the definition of $F^{(N)}$ and Theorem 1 together with
\reff{NFk} we obtain that higher order terms in the expansion of $F$
decay faster than $F^{(0)}$, and that their decay exponents satisfy
the inequality
\begin{equation}
N(F^{(n)}) \geq N(F^{(0)}) + n \lambda \qquad n\geq 0. \label{NFn}
\end{equation}

The functions \reff{metricfun} that appear in the Gowdy metric
\reff{gowdymetric}, expressed in terms of $x=X_1$ and $y=X_2$, are of
the form \reff{F}, or in the case of the function $A$, which is
defined by integral \reff{intA}, the integrand is of the form
\reff{F}. For the first three functions \reff{metricfun} the expansion
is straightforward and gives higher order corrections that are
globally defined functions on the $t\theta$ cylinder, whose decay
exponents satisfy \reff{NFn}.

The terms in the expansion of the function $A$ are
\begin{equation}
A^{(n)}(t,\theta) = A_0^{(n)} + \int_{t_0}^t
\alpha_t^{(n)}(t',\theta_0) dt' +
\int_{\theta_0}^\theta \alpha_\theta^{(n)}(t,\theta') d\theta',
\label{intAn}
\end{equation}
where $\alpha_t^{(n)}$ and $\alpha_\theta^{(n)}$ are terms in the
expansion of the functions $\alpha_t$ and $\alpha_\theta$ defined in
(11) and (12), and $A_0^{(n)}$ a suitable constant.

To determine the asymptotic behavior of the integrals note that, since
both $\alpha_t^{(n)}$ and $\alpha_\theta^{(n)}$ are $\xi$-expandable
functions we can integrate term by term. The time integral, when
evaluated at the upper limit $t$, gives a $\xi$-expandable function
with the decay exponent equal to the decay exponent of
$\alpha_t^{(n)}$. When evaluated at the lower limit $t_0$ it gives a
constant $k(t_0,\theta_0)$ that could spoil the decay properties of
$A^{(n)}$. To prevent that we choose $A_0^{(n)}=-k(t_0,\theta_0)$.

Using \reff{NFn} the theta integral can be shown to be
\begin{equation}
\left|
\int_{\theta_0}^\theta \alpha_\theta^{(n)}(t,\theta') d\theta'
\right|
\leq K^{(n)} \exp [- 2{ \;\rm inf} ( 1- c(\theta)) n t],
\end{equation}
for some constant $K^{(n)}>0$.

We have chosen a special path of integration for $A^{(n)}$, but it is
easy to show that $A^{(n)}$ is path independent. Using the functional
relation \reff{derivativealpha}
between $\alpha_\theta$ and $\alpha_t$ and reasoning
similar to that that led to \reff{allconstant} we obtain
\begin{equation}
\alpha_{\theta,t}^{(n)} = \alpha_{t,\theta}^{(n)} \label{pathindep}
\end{equation}
which guarantees the path independence. To have each $A^{(n)}$ defined
globally on the $t\theta$ cylinder they must be periodic in $\theta$,
with period $2\pi$,  which is equivalent to
\begin{equation}
C^{(n)}=\int_{0}^{2\pi} \alpha_\theta^{(n)}(t,\theta) d\theta =0.
\end{equation}
Equation \reff{pathindep} guarantees that $\dot C^{(n)}=0$, and we
have to impose the constraint at one time only.

The decay exponent of $\alpha_\theta^{(0)}$ is zero, which means that
$C^{(0)}=0$ will impose a constraint on the geodesic loop functions
$a,b,c$ and $d$. Evaluating $C^{(0)}$
and then going to the limit $t\rightarrow \infty$ we obtain
\begin{equation}
\int_{0}^{2\pi} c(\theta) \left[
\frac{a'(\theta)}{b(\theta))} + d'(\theta) \right] d\theta =0.
\label{constraint}
\end{equation}
For $n\geq1$ the decay exponents of $\alpha_\theta^{(n)}$ are,
according to \reff{NFn}, greater than 1, which together with the
constancy of $C^{(n)}$ implies that $C^{(n)}=0$, without the need to
impose any new constraints on $a,b,c$ and $d$.

So, we conclude that the expansion of the function $A$ is globally
defined provided the condition \reff{constraint} is satisfied by the
seed solution. Then, asymptotically for $t\rightarrow \infty$, the
zeroth order term in the expansion of $A$ is
\begin{equation}
A^{(0)}(t,\theta) \sim \frac{1}{4} \left[1-c^2(\theta)\right] t
-\frac{1}{2} \int_{\theta_0}^\theta
c(\vartheta)
\frac{a'(\vartheta)+b(\vartheta)d'(\vartheta)}{b(\vartheta)}
d\vartheta,
\end{equation}
and the higher order terms decay exponentially faster.

Therefore, we have shown  that for any geodesic loop
spacetime with metric $g^{(0)}$ whose geodesic loop functions $a$,
$b>0$, $c>0 $ and $d$ satisfy the constraint \reff{constraint} and
\begin{eqnarray}
c(\theta) < 1 \qquad {\rm or}  \label{speedless1} \\
c(\theta) {\rm arbitrary \quad and} \quad a(\theta) + b(\theta) =
{\rm
const.}, \label{polarization}
\end{eqnarray}
there exists a sequence $g^{(n)}$ of solutions of the perturbative
Gowdy equations, to arbitrarily high order $N$, such that,
asymptotically as $t \rightarrow \infty$,
\begin{equation}
\sum_{n=0}^N g^{(n)} \sim g^{(0)}
\end{equation}

The functions $a$, $b$, $c$, and $d$ satisfying \reff{constraint} and
\reff{speedless1} form an open subset in the set of all Gowdy initial
data. This gives us a hope that the exact Gowdy solutions from an open
subset $\cal G$ of all Gowdy solutions asymptotically approach
geodesic loop solutions. The geodesic loop approximation may not be
valid for all Gowdy spacetimes however. Consider the speed of the
exact Gowdy loops \reff{speed}; there may exist Gowdy spacetimes with
asymptotic speed $\lim_{t\rightarrow \infty} u=c>1$, and for such
spacetimes our perturbative expansion is not asymptotically dominated
by the zeroth order (the geodesic loop) term. In the case of $c>1$,
unless the  condition \reff{polarization} is satisfied, the higher
order terms in the expansion exponentially increase instead of
decrease compared to the zeroth order term, as $t \rightarrow \infty$.

All the rigorous results on Gowdy spacetimes obtained so far show that
the geodesic loop approximation is indeed asymptotically valid. In the
special case of the circular loop spacetimes, which because of
circular symmetry can't satisfy the condition \reff{polarization}, it
was rigorously proven \cite{chrusciel91} that the geodesic loop
approximation is asymptotically valid and that the asymptotic speed
$c$ is  always less than unity with all values between zero and unity
attained by some spacetime. Special condition \reff{polarization}
includes polarized Gowdy metrics which are rigorously known to
asymptotically approach geodesic loop spacetimes with arbitrary large
speed $c$ \cite{im90,cim90}. We will discuss these and some other
exact results in the last section.

For now, assume that our perturbative results are rigorously true,
i.e. that the geodesic loop approximation is valid for all spacetimes
with asymptotic speed $c<1$. The asymptotic behavior of curvature, as
$t \rightarrow \infty$, in Gowdy spacetimes can be calculated using
their corresponding  geodesic loop asymptotes. The asymptotic behavior
of the curvature in the geodesic loop spacetimes was analysed by
Mansfield \cite{mansfield89}. He showed that for $c\neq 1$ the
geodesic loop spacetime is curvature singular,  as $t \rightarrow
\infty$,  and
\begin{equation}
R^{\alpha \beta \gamma \delta} R_{\alpha \beta \gamma \delta}
\sim \frac{1}{4} e^{-4 f(\theta)} (c(\theta)^2 -1)^2 (c(\theta)^2 +3)
e^{(c(\theta)^2+3) t },
\end{equation}
where
\begin{equation}
f(\theta)=\int_{\theta_0}^\theta
c(\vartheta)
\frac{a'(\vartheta)+b(\vartheta)d'(\vartheta)}{b(\vartheta)}
d\vartheta.
\end{equation}
This means that the corresponding Gowdy spacetime with the asymptotic
speed $c<1$ is curvature singular, as $t \rightarrow \infty$, and
therefore inextendible beyond its maximal globally hyperbolic
development, in agreement with the SCC. Mansfield also showed that for
the special case of $c=1$ and $a+b=$const. the spacetime is not
curvature singular and is extendible.

\section{Concluding Remarks}
Chru\'{s}ciel and Moncrief  have derived some ``light cone'' and
``higher-order-energy'' estimates for Gowdy equations with a view
towards rigorously proving the geodesic loop asymptotic behavior
suggested by the perturbation calculations presented here
\cite{cmunpublished}. Several complications make such a proof
significantly more difficult here than for polarized case treated in
Refs.\cite{im90} and \cite{cim90}. Aside from the fact that now the
basic equations are non-linear there is also the complication that not
every geodesic loop can be realized as the asymptote of some exact
solution. The reason for this is that our perturbative calculations
require  $c(\theta)<1$ to be applicable, unless some special
non-generic condition such as $a+b=\rm const.$ is satisfied.

Nevertheless, it is quite conceivable that exact field equations
always force the asymptotic speed $c$ below unity (unless the
aforementioned special condition is satisfied) and thus force the
solutions to achieve the geodesic loop asymptotic behavior. An example
of this phenomenon was found by Chru\'{s}ciel and Moncrief who studied
exact circular loop solutions, the motion of which is governed by a
non-linear, second order {\em ordinary} differential equation (for the
``radius'' of the loop). The circularity is preserved by the exact
field equations and such solutions are prevented, by their circular
symmetry, from achieving the special condition which permits an
asymptotic speed greater than or equal to unity (unless the geodesic
loop behavior is violated). It was found, however, that every such
circular loop solution does indeed asymptotically approach a geodesic
loop (of speed less than unity) and that every value of asymptotic
speed (strictly less than unity) is in fact achieved by some exact
circular solution \cite{chrusciel91}.

The polarized solutions considered by Isenberg and Moncrief
\cite{im90} satisfy the special condition mentioned above
($a+b=$const.) and can achieve arbitrarily large asymptotic speeds
while still realizing the asymptotic geodesic loop behavior. Not only
is this asymptotic behavior universally satisfied by the polarized
solutions but also one  can rigorously justify computing the
asymptotic behavior of the Riemann tensor or (for sufficiently smooth
solutions) its covariant derivatives by means of computations made
purely within the geodesic loop approximation.

A further example of exact results which realize the geodesic loop
asymptotic behavior (this time with $c=1$) has been found by Mansfield
\cite{mansfield89} who transformed a class of exact analytic
solutions, derived using the Ernst formalism, back to the Gowdy
representation. This family of solutions (which is infinite
dimensional and generically non-polarized) turns out to be none other
than the set of ``generalized Taub-NUT'' spacetimes (restricted to the
Gowdy symmetry class) defined on $T^3 \times R$ which develop compact
Cauchy horizons instead of curvature singularity at the boundaries of
their maximal Cauchy development
\cite{moncrief82,moncrief84,moncrief87}. Much larger families of such
generalized Taub-NUT spacetimes (admitting generically only one
spacelike Killing field) are defined in Refs. \cite{moncrief82} and
\cite{moncrief84} and have been used in Ref.\cite{moncrief89} as
backgrounds for perturbative expansion. An infinite dimensional family
of analytic, curvature singular solutions having only one Killing
field (and defined on $S^2 \times S^1 \times R$) was constructed in
\cite{moncrief87} by applying suitably chosen Geroch transformation to
the generalized Taub-NUT solutions defined on $S^3 \times R$. One
hopes that these rigorously derived singular solutions will help one
to understand the asymptotic behaviors of the perturbative solutions
discussed in \cite{moncrief89}. At our present level of understanding
it seems not unreasonable to hope that the higher order perturbation
methods discussed here can be successfully applied to the study of
completely general (non-symmetric) cosmological solutions of
Einstein's equations near their singular boundaries.

\vskip 1in

{\bf \large Acknowledgments}

This work was supported in part by NSF Grants PHY-8903939 and
PHY-9201196 to Yale University.

%
%
%
% Phys. Rev. form

%
%
%
\end{document}